\documentclass[12pt]{article}
\usepackage{times}
\usepackage{color}
\usepackage{geometry}
\usepackage[utf8]{inputenc}
\usepackage{enumitem,amssymb}
\usepackage{ragged2e}
\newlist{thematic}{itemize}{8}
\setlist[thematic]{label=$\square$}
\usepackage{pifont}
\usepackage{natbib}
\setcitestyle{aysep={}}
\newcommand{\cmark}{\ding{51}}%
\newcommand{\done}{\rlap{$\square$}{\raisebox{2pt}{\large\hspace{1pt}\cmark}}%
\hspace{-2.5pt}}

\usepackage{graphicx}
\usepackage{parskip,multicol}
\usepackage{float}
\usepackage{wrapfig}

\usepackage{multicol}

\begin{document}
\raggedright
{\huge
Astro2020 Science White Paper \linebreak

Magnetism in the Brown Dwarf Regime} \linebreak
\normalsize

\noindent \textbf{Thematic Areas:} \hspace*{60pt} $\done$ Planetary Systems \hspace*{10pt} $\square$ Star and Planet Formation \hspace*{20pt}\linebreak
$\square$ Formation and Evolution of Compact Objects \hspace*{31pt} $\square$ Cosmology and Fundamental Physics \linebreak
  $\done$  Stars and Stellar Evolution \hspace*{1pt} $\square$ Resolved Stellar Populations and their Environments \hspace*{40pt} \linebreak
  $\square$    Galaxy Evolution   \hspace*{45pt} $\square$             Multi-Messenger Astronomy and Astrophysics \hspace*{65pt} \linebreak
  
\textbf{Principal Authors:}

\begin{multicols}{2}
Name:	Melodie M. Kao
 \linebreak						
Institution:  Arizona State University
 \linebreak
Email: mkao@asu.edu
 \linebreak
 
 Name:	J. Sebastian Pineda 
 \linebreak						
Institution:  University of Colorado Boulder
 \linebreak
Email: sebastian.pineda@lasp.colorado.edu
 \linebreak
\end{multicols}

\textbf{Co-authors:}\\
Peter Williams (Harvard), Rakesh Yadav (Harvard), Denis Shulyak (Georg-August U.), Joachim Saur (U.~Cologne), David J.\ Stevenson (Caltech), Sarah Schmidt (AIP), Adam Burgasser (UCSD), Gregg Hallinan (Caltech), Kelle Cruz (CUNY) \linebreak

\textbf{Abstract:}
\justifying{
A suite of discoveries in the last two decades demonstrate that we are now at a point where incorporating magnetic behavior is key for advancing our ability to characterize substellar and planetary systems. The next decade heralds the exciting maturation of the now-burgeoning field of brown dwarf magnetism, and investing now in brown dwarf magnetism will provide a key platform for exploring exoplanetary magnetism and habitability beyond the solar system. We anticipate significant discoveries including: the nature of substellar and planetary magnetic dynamos, the characterization of exo-aurora physics and brown dwarf magnetospheric environments, and the role of satellites in manifestations of substellar magnetic activity. These efforts will require significant new observational capabilities at radio and near infrared wavelengths, dedicated long-term monitoring programs, and committed support for the theoretical modeling efforts underpinning the physical processes of the magnetic phenomena. }
\pagebreak

\section{A revolution in brown dwarf magnetism: Recent advances}\label{recentAdv}
\vspace*{-0.6cm}

\justifying{Extensive observations in the last two decades have led to a leap in our understanding of magnetism in the brown dwarf regime.  Although weak chromospheric emission in L-dwarfs \citep{kirkpatrick2000} and cool effective temperatures \citep{mohanty2002} initially suggested that substellar objects would be magnetically inactive, a serendipitous first detection of flaring radio emission from a brown dwarf \citep{berger2001} demonstrated that strong magnetic activity extends to ultracool dwarfs ($\gtrsim$M7). Moreover, the unexpected 4--8 GHz radio flares departed dramatically from the tight G\"{u}del-Benz relationship linking radio and X-ray luminosities in magnetically active stars \citep{williams2014}. In short, substellar versus stellar magnetic activity seemed different. 
\vspace{-3pt}

The subsequent identification of the flaring component of ultracool dwarf radio emission as coherent electron cyclotron maser emission (ECM) confirmed that ultracool dwarfs can host kilogauss (kG) magnetic fields \citep{hallinan2006, hallinan2008}.  Soon thereafter, the discovery of the first T dwarf ($\sim$890 K) at 4--8 GHz \citep{route2012} demonstrated that magnetic behavior persists well beyond the observed drop-off in chromospheric emission/white-light flaring seen across early-to-mid L spectral types \citep{schmidt2015,pineda2016,gizis2017}.

\vspace{-3pt}
Concurrently, Zeeman broadening and Zeeman Doppler Imaging (ZDI) techniques revealed that magnetic topologies of fully convective objects like brown dwarfs fall into two main categories: either predominantly dipolar and globally kG fields or weaker global fields with localized multipolar kG regions \citep{morin2010,morin2011}.  Surface-averaged field strengths were initially thought to saturate at fast rotation rates \citep{reinersBasriBrowning2009}, but recent Zeeman broadening innovations demonstrate that dipolar field strengths ($\gtrsim$4 kG) can exceed saturation levels in the strongest mulipolar objects \citep{shulyak2017}. These stronger fields are in accord with observed kG fields in rapidly rotating brown dwarfs \citep{kao2018} and imply that dipolar magnetic topologies may be important for strong ECM radio activity \citep{pineda2017,pineda2018}.

\vspace{-0.6cm}
{\color{blue}\subsection{A new paradigm shift in substellar magnetic activity: Exo-aurorae} \label{sec:intro_paradigm}}
\vspace*{-0.5cm}
Our understanding of substellar magnetic activity is now undergoing a new paradigm shift.  The above discoveries together with studies linking ECM emission to H$\alpha$ and infrared variability \citep{harding2013, hallinan2015, kao2016} suggested that brown dwarf ECM emission is an auroral signature rather than flaring magnetic reconnection \citep{hallinan2015}. The auroral paradigm led to the first detection of radio emission from a planetary-mass object \citep{kao2016, kao2018}, which revealed order-of-magnitude stronger magnetic fields than predicted by planetary dynamo models and provided an inroad to studying exoplanetary magnetism.

\vspace{-3pt}

\textbf{Together, these detections are inspiring a wide breadth of investigations into the effects of magnetism on substellar objects, their observable properties, and the underlying physics governing these effects from stars to planets.}  We are studying the impact of auroral electron beams on brown dwarf atmospheres \citep{pineda2017} and the nature of the electrodynamic engine driving ECM emission \citep{schrijver2009,nichols2012, turnpenney2017}, testing substellar magnetic dynamo mechanisms \citep{kao2016, kao2018}, modeling the magnetic topology in ECM source regions \citep{yu2011, lynch2015, leto2016}, and conducting multiwavelength searches for magnetic star-planet interactions \citep{hallinan2013,turnpenney2018,pineda2018}. With these recent advances, the next decade is poised to address many outstanding questions in brown dwarf magnetism and enter into an era of characterizing the full magnetic ecosystems of brown dwarfs and their close cousins, exoplanets.}

\section{This decade: Characterizing substellar magnetic ecosystems}
\vspace*{-0.5cm}
{\color{blue}\subsection{Brown dwarf dynamos and magnetic fields} \label{sec:bfield_obs}}
\vspace*{-0.5cm}

Two key goals in substellar magnetism are characterizing (1) magnetic fields and (2) the engines that generate them. Dynamo mechanisms that rely on convecting fluids are expected to be ubiquitous in brown dwarfs and exoplanets, making brown dwarfs excellent tests of exoplanetary magnetism. A needed direction of growth will be the measurement of magnetic field properties (strength, topologies, cycles) spanning a wide range of brown dwarf properties (masses, temperatures, rotation rates, and ages).  Together, these data will identify empirical scaling relationships between magnetic properties and object fundamental properties.  Simultaneously, improved dynamo simulations are needed to illuminate the underpinning physics of these relationships.

\vspace{-0.5cm}
\subsubsection{What determines the properties of magnetic dynamos in the substellar regime?} 
\vspace{-0.5cm}

One frontier of testing planetary and substellar dynamo mechanisms will be with measurements of magnetic fields at young ages ($\lesssim$200 Myr) and planetary masses ($\lesssim$12 M$_{\mathrm{J}}$) using young brown dwarfs identified in moving groups \citep[e.g.,][]{gagne2017}. These objects are laboratories for assessing magnetic evolution and provide a benchmark for field generation in planetary-like dynamos, which impact exoplanet atmospheric evolution and habitability \citep[e.g.,][]{khodachenko2007, vidotto2013}. Additionally, through measurements of their magnetic fields, the newly discovered Y dwarfs \citep{cushing2011} with effective temperatures of $\sim$350 K  will be the frontier for probing dynamo physics occurring at planetary temperatures.
\begin{wrapfigure}{R}{0.45\textwidth}
\vspace{-0.6cm}
\begin{center}
    \includegraphics[width=0.45\textwidth]{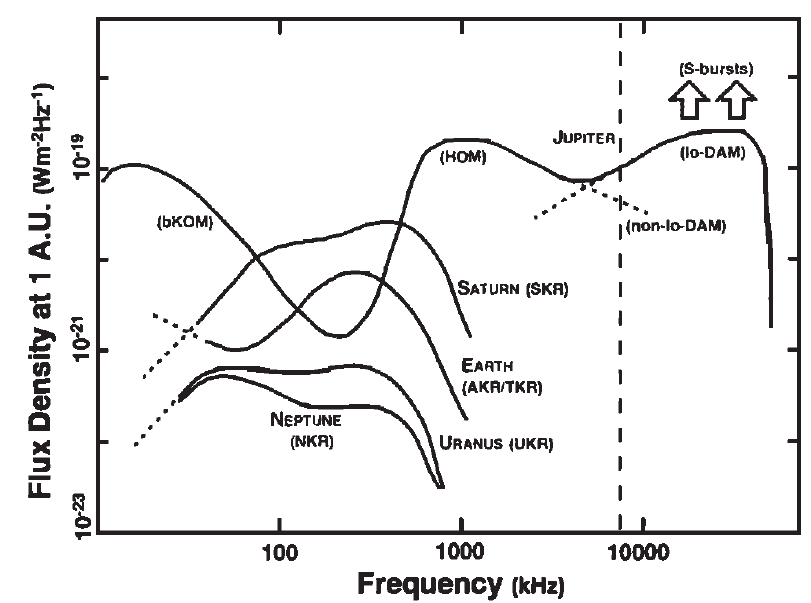}
    \vspace{-0.9cm}
\caption{\small{Detecting auroral radio emission drop-offs, such as these for solar system planets \citep{zarka1998}, places strong constraints on surface magnetic field strengths and will be key for deriving empirical dynamo scaling relationships.}}\label{fig:cutoff}
\end{center}
\vspace{-0.3cm}
\end{wrapfigure}

\vspace{-3pt}
Measuring dynamo scaling relationships will require more accurate constraints on the magnetic field strengths of brown dwarfs. A key area of progress will be radio observations at higher frequencies to measure the distinct drop-off in the emission of radio aurorae (see Fig.~\ref{fig:cutoff}) that maps directly to the strongest local magnetic fields of the source region.  Only five brown dwarfs have measurements between 8--12 GHz (2.9--4.3~kG), and just one has a measurement at 12--18 GHz (4.3--6.4~kG), with no definitive detections of the auroral cutoff yet \citep{kao2018}. Additionally, Zeeman broadening measurements (see \S \ref{sec:mag_env}) will provide complementary information about surface-averaged fields for comparison to the localized radio measurements.

\vspace{-0.5cm}
\subsubsection{Do brown dwarfs host magnetic cycles?} 
\vspace{-0.5cm}
Preliminary evidence of magnetic polarity reversals in radio brown dwarfs \citep{route2016} opened the possibility of such cycles extending to the substellar regime.  Investigating this possibility calls for long-term monitoring of brown dwarf magnetic fields. Multi-epoch Zeeman broadening and ZDI measurements can reveal time variability in large- vs.\ small-scale magnetic fields (see \S \ref{sec:mag_env}). Combining these data with models of ECM emission \citep[e.g.,][]{lynch2015,leto2016} will further probe the variable field and plasma conditions at source regions.

\vspace*{-0.5cm}

{\color{blue}\subsection{Magnetospheric environments around brown dwarfs}}
\vspace*{-0.5cm}
Despite significant progress in measuring the magnetic field strengths of brown dwarfs, little is known about their magnetospheric topologies or environments. ECM radio measurements cannot currently assess global properties of magnetic fields \citep{kao2018}, and attempts to infer topologies from the ECM emission of individual objects yield contradictory results \citep[e.g.,][]{yu2011, kuznetsov2012,lynch2015}. In contrast, Zeeman broadening and ZDI can yield topological information, but neither currently extend to L or later spectral types ($\lesssim$2000 K). 
\vspace*{-0.5cm}  
\subsubsection{What magnetic topologies can generate aurora?} \label{sec:mag_env}
\vspace*{-0.5cm}  
A necessary direction will be extending these Zeeman techniques to L and later type brown dwarfs. This will require new laboratory measurements and/or detailed theoretical treatments of Land\'{e} factors to identify and model the spectropolarimetic signatures of magnetically sensitive features (e.g., CrH/FeH; \citealt{kuzmychov2017}).  Combining ZDI magnetic maps of brown dwarfs with Very Long Baseline Interferometry (VLBI) imaging of ECM emission will help identify whether the latter can originate from large-scale and/or small-scale magnetic field components \citep[e.g.][]{berdyugina2017}, and assess the role of field topology in the transition from coronal to auroral activity. 

\vspace*{-0.5cm} 

\subsubsection{Do brown dwarfs have radiation belts or Io plasma torus analogs?} \label{sec:radBelt}
\vspace*{-0.5cm}  

An intriguing question raised by recent detections of radio emission in brown dwarfs is the source of the non-pulsing (quiescent) radio emission that accompanies all pulsing radio aurora. 95 GHz observations of the brown dwarf TVLM 513-46546 confirm a spectral index consistent with gyrosynchrotron radiation \citep{williams2015b}. Next generation VLBI radio facilities will allow for resolved imaging of brown dwarf magnetospheres to search for the presence of plasma structures (for example, extrasolar radiation belt or Io plasma torus analogs; \citealt{bagenal1981}) in the magnetospheres of brown dwarfs as one possible source of the quiescent radio component.

\vspace*{-0.5cm} 
{\color{blue}\subsection{Physical processes driving multi-wavelength auroral emissions}\label{sec:aurora}}
\vspace*{-0.5cm} 
\subsubsection{What powers the electrodynamic engine?}\label{sec:engine}
\vspace*{-0.5cm}
The identification of brown dwarf radio pulses as aurorae has sparked new questions regarding how they are sustained and powered; chiefly, what is the nature of the underlying electrodynamic engine?  Current theories include a satellite flux-tube interaction (see \S~\ref{sec:moons}), an internally driven magnetospheric engine (co-rotation breakdown of a plasma disk), or reconnection with the ISM \citep{nichols2012, turnpenney2017, pineda2018}. Understanding the observed relationship between the physical properties of the brown dwarfs and the strength and/or presence of auroral activity is key in distinguishing these models and how they drive the observed occurrence rate of brown dwarf aurorae \citep[$\lesssim10\%$;][]{routeWolszczan2016b, lynch2016}. 

\vspace{-3pt}
To these ends we need new radio monitoring programs to expand the current sample of radio brown dwarfs \citep[$<$30;][]{pineda2017,kao2018}. Such programs will need to identify or rule out periodic pulsations of $\sim$10 $\mu$Jy on minute timescales, and extend to the slower rotators (P$\gtrsim$10 hr) to test the dependence of the observed radio emission strength on rotation period in comparison to model predictions \citep[e.g.,][]{nichols2012}.  Also, new measurements of substellar magnetic field strengths (see \S~\ref{sec:bfield_obs}) from active and inactive objects will test model predictions for pulse luminosity as a function of magnetic field strength \citep[e.g.,][]{turnpenney2017}.

\vspace*{-0.5cm} 
\subsubsection{What is the impact of the auroral electron beam?}
\vspace*{-0.5cm}
The observed ECM radio emission requires an energetic electron beam \citep{treumann2006}. These auroral electrons both generate the radio emission and produce a cascade of emissions from the UV to the IR as their energy is deposited into the atmosphere \citep{badman2015,pineda2017}. New modeling efforts are needed to understand what impact the electron beams have on the brown dwarf atmosphere/ionosphere to compare to current observations \citep[e.g.,][]{hallinan2015,saur2018}.  A detailed accounting of the energy budget will provide constraints on the electron energy distribution, the strength of the auroral electrodynamic engine, and the efficiency of the ECM proces -- all significant unknowns in the theory of brown dwarf aurorae. 

\vspace*{-0.5cm} 
{\color{blue}\subsection{Brown dwarf satellites and magnetic interactions}\label{sec:moons}}
\vspace*{-0.5cm} 
\subsubsection{Do brown dwarf satellites power aurorae?}
\vspace*{-0.5cm}

In the next decade we can confirm or rule-out brown dwarf satellites (e.g., moons and exoplanets) as agents in the auroral electrodynamic engine (see \S \ref{sec:engine}).  In satellite flux-tube interactions, closely orbiting planets generate current systems by perturbing the rotating brown dwarf magnetosphere \citep{zarka2007,saur2013}. Satellites may also serve as the source for the magnetospheric plasma, as Io does in the Jovian magnetosphere, which powers auroral currents through co-rotation breakdown \citep{hill2001,cowleyBunce2001}. No satellite has been confirmed around any radio brown dwarf; however, the M8 TRAPPIST-1 system  suggests that planet formation persists toward the smallest stars and substellar objects \citep{gillon2017}. Sensitive exoplanet searches in this regime will lay the ground work for statistical comparisons of planet and auroral occurrence rates, and follow-up observations with extended monitoring could reveal conclusive evidence of satellite interactions \citep{fischer2019}.

\vspace*{-0.5cm} 
\subsubsection{What are the magnetic properties of exoplanets?}
\vspace*{-0.5cm}

\begin{wrapfigure}{r}{0.5\textwidth}
\vspace{-0.7cm}
\begin{center}
\includegraphics[width=0.5\textwidth]{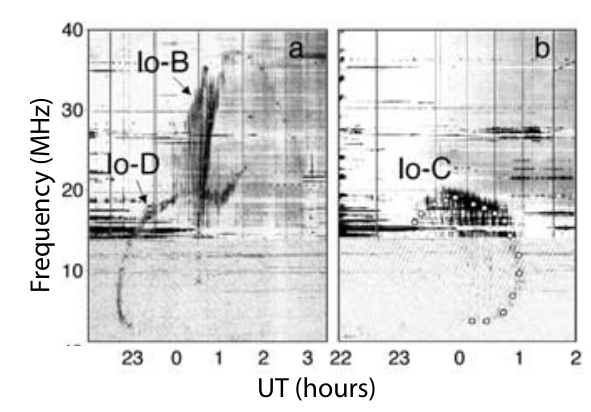}
\vspace{-1cm}
\caption{\small{Radio interferometers that yield exquisite dynamic spectra of characteristic arcs from host-satellite auroral interactions like these for Io/Jupiter \citep{hess2008} will reveal satellite properties. 
}}
\label{fig:ioArcs}
\vspace{-0.5cm}
\end{center}
\end{wrapfigure}

Satellite-driven aurorae can give insights into the natures of the satellites themselves.  The strength of the auroral emission depends on the size of the satellite obstacle, either its ionosphere or magnetosphere \citep{wannawichian2010,saur2013}. Additionally, magnetospheric satellite interactions show uniquely characteristic arcs in the dynamic spectra (time- and frequency- dependent flux densities) of radio aurora (see Fig. \ref{fig:ioArcs}).  Detailed models of these radio arcs can yield information about the orbits, rotation, and magnetic fields of satellites \citep[e.g.,][]{hess2008, hess2011, hess2012}, and similar models are being developed for future detections of exoplanet radio emission \citep{louis2019}. These discoveries will be opportunities to apply lessons from the current \textit{Juno} mission to Jupiter \citep{bagenal2017}, testing how auroral physics changes in different environments as compared to the solar system. Adapting these models to brown dwarfs together with investing in next generation radio interferometry facilities that can yield exquisite dynamic spectra of radio brown dwarfs will give us a window into exoplanetary magnetism.

\vspace*{-0.5cm}  

{\color{blue}\subsection{Influence of magnetic fields on atmospheres and structure}}
\vspace*{-0.5cm}  
\subsubsection{How do atmospheres change from stellar to planetary regimes?}
\vspace*{-0.5cm}

It is now clear that magnetism transitions from stellar-like coronae in low-mass stars to planet-like aurorae in the coolest brown dwarfs \citep[][]{kao2016,pineda2017}. This change also manifests in observable atmospheric features. Whereas stars possess starspot structures \citep[e.g.,][]{barnes2015} and cool brown dwarfs show cloud signatures \citep{radigan2014a}, transition region behavior remains uncertain for objects in between \citep{gizis2015}. Disentangling these effects across the substellar regime will be possible through dedicated spectrophotometric monitoring to reveal both the wavelength dependence of these features and their characteristic time variability.

\vspace*{-0.5cm}  
\subsubsection{How do magnetic fields impact stellar structure?}
\vspace*{-0.5cm}  

The observed magnetic fields of brown dwarfs suggest that they may have an important impact on substellar structure \citep[e.g.,][]{stassun2006}, with the most significant discrepancies between models and observed properties seen in the lowest mass objects \citep{stassun2012,kesseli2018}. 
Inflated radii \citep[e.g.,][]{batygin2010} and mis-characterized temperatures or ages will distort model-based inferences, impacting the assumed properties of planets found around brown dwarfs. Strong magnetic fields must be incorporated into next generation structure and evolution models of substellar objects. Moreover, extensive all-sky photometric monitoring will yield a much larger sample of eclipsing binaries across a range of ages to serve as model benchmarks.

\vspace*{-0.6cm} 
\section{Recommendations}
\vspace*{-0.5cm}  

Discoveries in the last two decades have demonstrated that understanding substellar magnetism will be a key factor in illuminating the characteristics of substellar and planetary systems. By capitalizing on a newly emergent paradigm of auroral magnetic activity in brown dwarfs, the decade 2020--2030 is poised to revolutionize our understanding of the physics underpinning the manifestation of substellar magnetic fields and aurorae, characterize substellar magnetospheric environments, and assess the broad impact of magnetism on our understanding of exo-moons and substellar atmospheres. Therefore, we recommend the following investments:

 \vspace*{-0.1cm}
\begin{itemize}
    \item Achieving 1--20 GHz radio interferometric sensitivities of $\sim$0.1 $\mu$Jy in 2 hr exposures to increase discovery parameter space with deeper searches and detailed characterization
        \vspace*{-0.15cm}
    \item VLBI capabilities in next generation radio facilities for magnetospheric imaging
         \vspace*{-0.15cm}
    \item Developing radio all-sky survey camera technology with sufficient baseline coverage to eliminate computationally expensive deconvolutions, for long-term magnetic monitoring 
         \vspace*{-0.15cm}
    \item R $\gtrsim$ 10$^{5}$ spectropolarimeters ($\lambda\gtrsim$ 1 $\mu$m) reaching SNR$\sim$100 in tens of min.\ for $J$$\sim$10--15
        \vspace*{-0.15cm}
    \item Improved experimental/theoretical treatments of magnetically sensitive lines in the IR 
  \vspace*{-0.15cm}
    \item Spectroscopic (photometric/polarimetric) short- and long-term monitoring of brown dwarfs
    \vspace*{-0.15cm}
    \item Developing full-body dynamo simulations that include outer turbulent layers
     \vspace*{-0.15cm}
    \item Building models of auroral physics in brown dwarf environments
         \vspace*{-0.15cm}
    \item Adopting a PI funding model for flagship radio facilities to address data processing and access bottlenecks from high computational and data storage costs of radio observations 
\end{itemize}

\pagebreak
\section*{References}

\setlength{\bibsep}{-1pt plus 0.25ex}
\renewcommand\refname{\vskip -8mm}
\begin{multicols}{2}{\footnotesize
\bibliography{BIB_decadal2020.bib}
\bibliographystyle{apj_short_prop.bst} 

}
\end{multicols}

\end{document}